\documentclass[aps,showpacs,prl,twocolumn,amsmath,amsfonts,amssymb,superscriptaddress]{revtex4}

\usepackage{graphicx}

\let\s=\sigma   
   \let\G=\Gamma
\let\D=\Delta   
   
   \let\io=\infty

\def\to{\rightarrow} \def\la{\left\langle} \def\ra{\right\rangle}

\newcommand{\beq}{\begin{equation}} \newcommand{\eeq}{\end{equation}}

\begin{document}

\title{First-order transitions and the performance of quantum algorithms in
random optimization problems}

\author{Thomas J\"org} \affiliation{LPTENS, CNRS UMR 8549, associ\'ee
  \`a l'UPMC Paris 06, 24 Rue Lhomond, 75005 Paris, France.}

\author{Florent Krzakala} \affiliation{ ESPCI ParisTech, CNRS UMR 7083
  Gulliver, 10 rue Vauquelin, 75005 Paris, France}
\affiliation{T-Division and Center for Nonlinear Studies, Los Alamos
  National Laboratory, NM 87545 USA}

\author{Guilhem Semerjian} \affiliation{LPTENS, CNRS UMR 8549,
  associ\'ee \`a l'UPMC Paris 06, 24 Rue Lhomond, 75005 Paris,
  France.}

\author{Francesco Zamponi} \affiliation{Princeton Center for
  Theoretical Science, Princeton University, Princeton, NJ 08544, USA}
\affiliation{LPTENS, CNRS UMR 8549, associ\'ee \`a l'UPMC Paris 06, 24
  Rue Lhomond, 75005 Paris, France.}


\begin{abstract}
  We present a study of the phase diagram of a random
  optimization problem in presence of quantum fluctuations.
  Our main result is the characterization of the nature of the
  phase transition, which we find to be 
  a first-order quantum phase transition.
  We provide evidence that the gap
  vanishes exponentially with the system size at the transition. This
  indicates that the Quantum Adiabatic Algorithm requires a time growing
  exponentially with system size to find the ground state of this problem.
\end{abstract}

\maketitle

Theoretical research on quantum computing is motivated by the exciting
perspective of computers that take intrinsically advantage of the laws
of quantum mechanics.  Besides the great effort of research towards
the physical realization of these devices, a lot of activity has been
devoted to the development of algorithms that could use quantum
properties to achieve a faster velocity in performing computational
tasks with respect to classical devices.  A typical problem that is
encountered in almost all branches of science is that of optimizing
irregularly shaped cost functions: the Quantum Adiabatic Algorithm
(QAA) \cite{annealing,adiabatic} is in principle able to tackle such
problems in a universal way.  Suppose one wishes to find the ground
state of a Hamiltonian $H_P$ acting on $N$ qubits.  To perform the QAA
one considers a simpler Hamiltonian $H_Q$, such that the quantum
computer can be easily initialized in its ground state. If one slowly
interpolates the Hamiltonian $H(t)$ of the quantum computer from $H_Q$
to $H_P$, the adiabatic theorem ensures that, with high enough
probability, the system will remain at all times in the ground state
of the interpolating Hamiltonian.  Hence, at the end of the evolution,
it will be in the ground state of $H_P$ and the original problem will
be solved.  The crucial question is of course, how slow the evolution
should be in the thermodynamic limit $N\!\to\!\io$.  Quite generally, the
adiabaticity condition requires the rate of change of $H(t)$ to be
smaller than the (squared) gap between the ground state and the first
excited state of $H(t)$.  Hence, the time needed to ensure
adiabaticity will diverge in the thermodynamic limit whenever a
quantum phase transition, at which the gap is expected to
vanish~\cite{Sachdev}, is encountered during the interpolation between
$H_Q$ and $H_P$.  It is well established that the gap vanishes {\it at
  least polynomially} in $N$ at a quantum second-order critical
point~\cite{Sachdev} (except in some cases in presence of
disorder~\cite{Fisher}), while it vanishes {\it exponentially} in $N$
at a first-order phase transition~\cite{QREM,First,FirstY}.  First-order
phase transitions are thus particularly dangerous for the QAA.

The formal computational complexity theory classifies the difficulty
of a problem according to a worst-case criterion. It might, however,
well be that ``most'' of the problems in a given class are easy, even
though a few atypical instances are very difficult. To give a precise
content to this notion of typicality the research has turned to the
study of random instances, defining a probability distribution on the
space of instances. Statistical mechanics tools have provided a very
detailed and intricate picture of the properties of the configuration
space of such typical problem Hamiltonians $H_P$~\cite{CSP_classical}.
Random instances were also used to benchmark the performance of the
QAA, and early results generated considerable excitement by reporting
polynomial scaling of the minimum gap for sizes up to $N\sim
100$~\cite{adiabatic,Second}. However, some evidence of the presence
of first-order phase transitions has been recently
reported~\cite{QREM,First,FirstY}, which is natural from the point of view of
mean-field quantum spin glasses~\cite{FirstMF}.  These studies rely on
numerical investigations of small systems ($N \lesssim 256$) and/or on
perturbation theory close to $H_P$, hence they should be completed by
a non-perturbative analytic treatment in the thermodynamic limit.

\begin{figure}
  \includegraphics[width=.85\columnwidth]{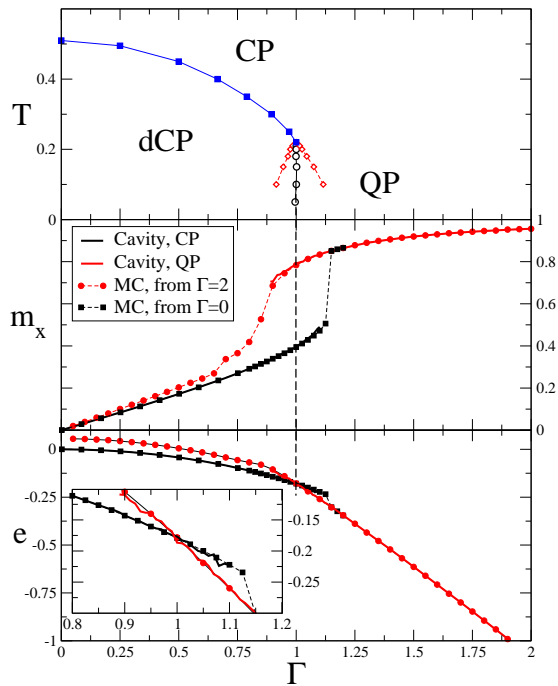}
    \vskip-10pt
  \caption{ (Top) Phase diagram of Eq.~(\ref{H}) for $c\!=\!3$. Open
    symbols are results of the RS calculation: first-order transition
    line separating the CP and QP (circles), with the corresponding
    spinodals (diamonds). Full symbols are the result of the 1RSB
    calculation: squares, clustering transition separating the CP and
    dCP.  (Bottom) Energy and $m_x$ as a function of $\G$ for
    temperature $T\!=\!0.05$. MC data for a sample with $N\!=\!2049$.}
  \label{fig:1}
\end{figure}

\begin{figure}
  \includegraphics[width=.85\columnwidth]{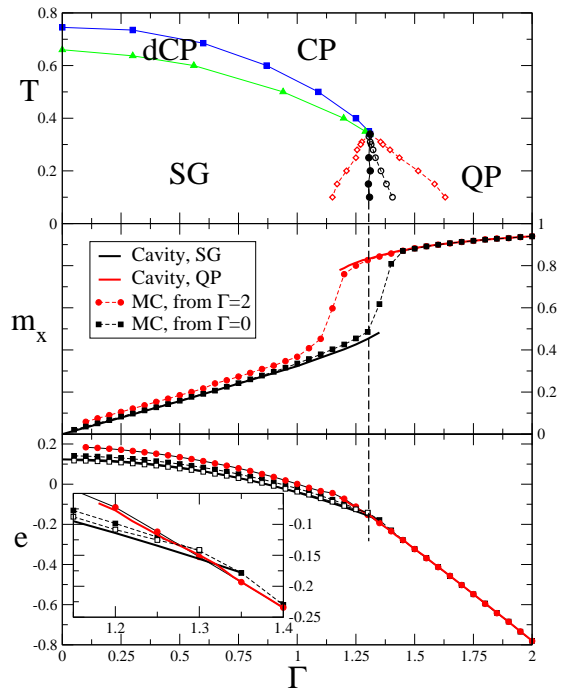}
    \vskip-10pt
  \caption{ (Top) Phase diagram of Eq.~(\ref{H}) for $c=4$. Symbols as
    in Fig.~\ref{fig:1}, with the addition of the spin-glass
    transition line (full triangles); the correct first-order
    transition line is obtained from the 1RSB calculation (full
    circles).  (Bottom) Energy and $m_x$ as a function of $\G$ for
    temperature $T=0.05$.  MC data for $N=120$ and averaged over 20
    samples (full symbols) and extrapolated in $1/N$ to the $N\!\to\!\io$ limit (open symbols).  
    Black curve, starting from the classical ground state
    found using an exact MAXSAT solver \cite{max1}. Red curve,
    starting from the QP.  }
  \label{fig:2}
\end{figure}

This is what we achieve in this Letter, reporting the results of 
the first analytical
study of a {\it random, quantum, finite-connectivity optimization
  problem} (namely, random regular 3-XORSAT in a transverse field),
that is believed to be largely representative of the generic behavior
of these problems. We compute the complete phase diagram of the model
{\it in the thermodynamic limit $N \!\to\! \io$ and for a uniformly random
  distribution of instances}, as a function of temperature and
transverse field. This is possible thanks to the
{\it quantum cavity method}, recently introduced in~\cite{QCM} and further
developed in~\cite{QCM2} (see~\cite{SQCM} for related work), that allows 
to solve exactly 
these problems by generalizing to the quantum case
the method developed for classical models~\cite{MePa}. Our main result
is the occurrence of a first-order phase transition at zero
temperature as a function of the transverse field. We corroborate the
analytical results with Exact Diagonalization and Quantum Monte Carlo
data. We provide evidence that the gap vanishes
exponentially in the size of the system at the transition.  These
results strongly suggest that the QAA requires an exponentially large
time in $N$ to find the ground state of this problem.

\begin{figure}
  \includegraphics[width=.85\columnwidth]{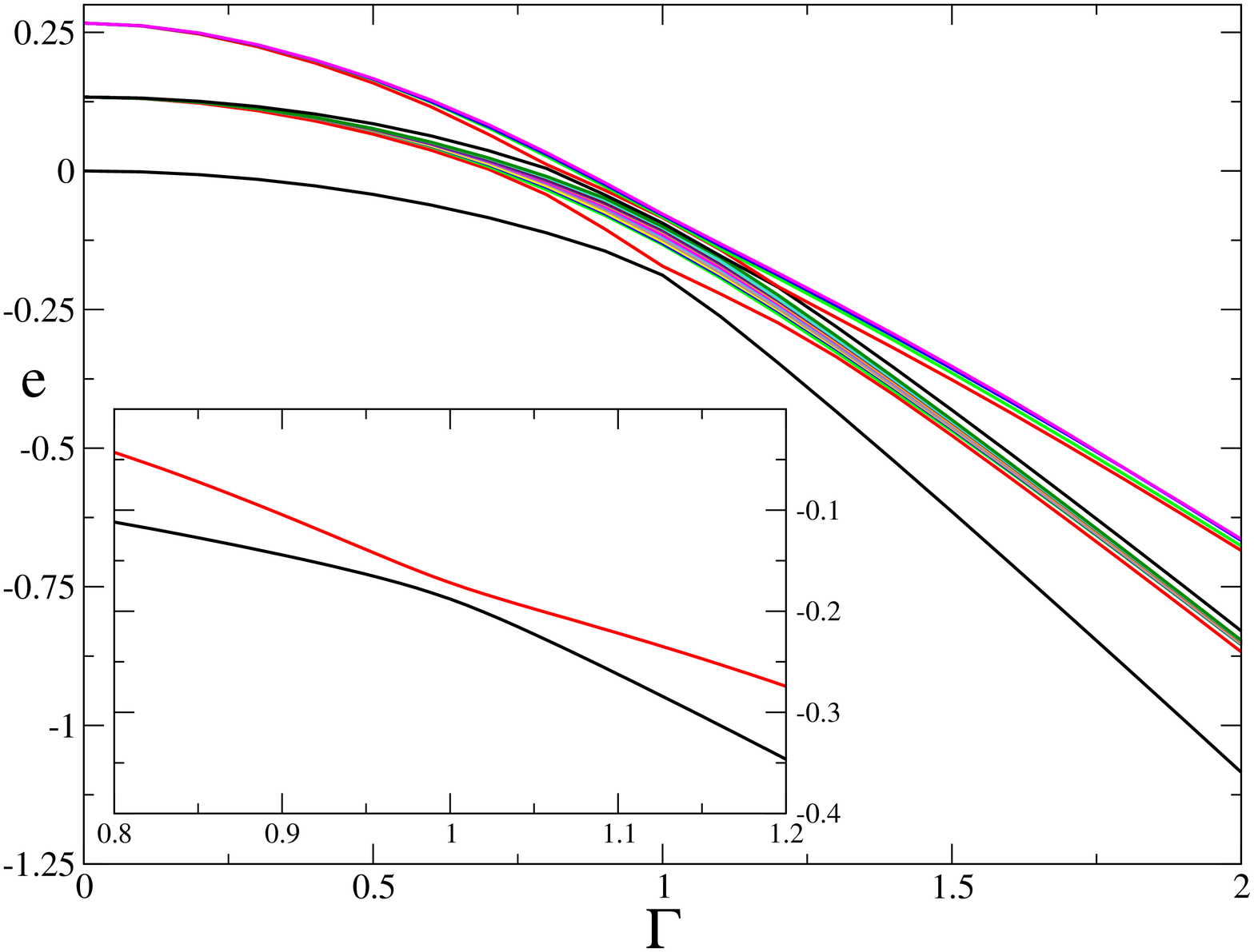}
    \vskip-10pt
  \caption{ Lowest energy levels from exact diagonalization of a USA
    instance with $c=3$ and $N=15$. In the inset the region close to
    the phase transition is magnified.
     }
      \label{fig:3}
\end{figure}

\paragraph*{Methods.---} 
We focus on the random $c$-regular $3$-XORSAT
problem~\cite{classic1RSB}, that in spin language is defined by the
Hamiltonian
\beq
\label{H} H = H_P + H_Q = \sum_{a=1}^M (1- J_a
\s_{i^a_1}^z \s_{i^a_2}^z \s_{i^a_3}^z) - \G \sum_{i=1}^N \s^x_i \ .
\eeq 
Here, $J_a = \pm 1$ with equal probability. The $3$ spins $i^a_1,
i^a_2, i^a_3$ involved in clauses $a = 1, \cdots, M = N c/3$ are
chosen uniformly at random among all possible choices such that each
spin enters {\it exactly} in $c$ clauses.  In the classical limit
$\G=0$, a given instance of the system is called {\it satisfiable}
(SAT) if there is a ground state of zero energy, UNSAT otherwise.

The thermodynamic properties of the model in the thermodynamic limit
$N\!\to\!\io$, and on average over the disorder, can be obtained by means
of the {\it cavity method}~\cite{MePa}. For quantum models, the cavity
method allows to reduce the solution of the model to the problem of
finding the fixed point of a functional equation for the local spin
effective actions~\cite{QCM,QCM2} (this functional equation being solved
numerically as in~\cite{QCM2} with a population dynamics algorithm~\cite{MePa}).  In
presence of one single pure state, the method discussed
in~\cite{QCM,QCM2}, that goes under the name of {\it replica symmetric}
(RS), is enough to obtain the correct solution. However, in order to
describe the low-temperature glassy phase, which is characterized by a
large number of pure states, one has to introduce a generalization of
the RS cavity method that goes under the name of {\it one-step replica
  symmetry breaking} (1RSB)~\cite{MePa}. This generalization, that we
introduce here for the first time in the context of quantum diluted
models, proceeds along the line of the classical computation (see
\cite{classic1RSB} for the solution of (\ref{H}) at $\G=0$) using as
variables the imaginary-time spin trajectories as detailed
in~\cite{QCM2}.  Additionally, for finite $N$, we performed Exact
Diagonalization (ED) using the Ritz functional method~\cite{Rmet}, and
Quantum Monte Carlo (QMC) simulations using the heat-bath algorithm
introduced in~\cite{QCM2}.

\paragraph*{Results of the cavity method.---}
In the classical limit  \cite{classic1RSB} when $\G\!=\!0$,
the model is SAT (with a probability going to 1 as $N \to \infty$) for
$c<3$, UNSAT for $c>3$, while in the marginal case $c=3$ it is SAT
with finite probability.  Let us begin the description of our results
with the simpler case $c=3$.  The RS computation predicts, at low
enough temperature $T \lesssim 0.24$, a first-order transition between
two different paramagnetic ($m_z = \la \s^z_i \ra =0$) phases: the
{\it Classical Paramagnet} (CP) characterized by a small value of
transverse magnetization $m_x = \la \s_i^x\ra$, and the {\it Quantum
  Paramagnet} (QP) that has a larger value of $m_x$.  This transition
and the corresponding spinodals are shown in the $(\G,T)$ phase
diagram of the top panel in Fig.~\ref{fig:1}; the transition is found
around $\G_{\rm c}(T)\approx 1.0$ for all values of $T \lesssim
0.24$. We also report in the bottom panel of Fig.~\ref{fig:1} the
cavity method predictions for $m_x$ and the
energy density $e=\la H \ra/N$ at very low temperature.  The outcome
of the 1RSB computation is twofold: it confirms that the RS
computation of the thermodynamic observables is in this case correct
in the whole phase diagram $(\G,T)$, in particular they are singular
only on the RS transition line. Moreover it unveils that, for low
enough values of $T$ and $\G$ the CP phase is actually a ``dynamical
CP'', in technical terms a 1RSB phase with Parisi breaking parameter
$x$ equal to 1, where an exponential number of pure states
coexist. The attribute dynamical, taken from the literature on
classical mean-field spin glasses and optimization
problems~\cite{CSP_classical}, emphasizes that equilibrium
thermodynamic properties are unaffected as one crosses the line
between CP and dCP (also plotted in Fig.~\ref{fig:1}).

We turn now to the $c=4$ case (which is representative of the behavior 
for any $c > 3$), for which the results are displayed in
a similar fashion on Fig.~\ref{fig:2}. It has a richer phenomenology
very similar to the one of fully-connected mean-field
models~\cite{FirstMF}.
Indeed the dCP undergoes a thermodynamically
second-order phase transition to a true {\it Spin-Glass} (SG) phase
(with a sub-exponential number of pure states and $x<1$). At low
enough temperature the thermodynamic transition becomes first order,
between the 1RSB SG and the QP. For this reason, the RS computation
gives a wrong result for the first-order transition line, see top
panel of Fig.~\ref{fig:2}.  In both cases we conclude on the existence
of a {\it first-order quantum phase transition} at $\G = \G_{\rm c}$
and zero temperature, separating the dCP (for $c=3$) or the SG (for
$c=4$) from the QP. The transition extends in a line $\G_{\rm c}(T)$ 
at low enough temperature, which is almost independent of $T$.

\paragraph*{Numerical investigations.---}
A first instructive example of the relevance of this transition is
found by comparing the cavity results with QMC (Figs.~\ref{fig:1},
\ref{fig:2} lower panel).  
We first run a QMC starting from
the classical ground state at $\G=0$ and slowly increasing $\G$. For
$c=3$, instances have a finite probability of being SAT, and otherwise
have an energy of order $1/N$ (see below
and~\cite{classic1RSB}): since SAT instances can be solved in
polynomial time using the Gauss elimination algorithm, we can limit
ourselves to the study of this simple case, and we can then run the
QMC for very large sizes ($N=2049$). For $c=4$, the problem is
typically UNSAT~\cite{classic1RSB}, finding the ground state is very
hard (actually, NP-hard), and we are limited to much smaller sizes ($N \leq 120$); yet a good
extrapolation in $1/N$ to the thermodynamic limit is possible. 
In both cases we find that the QMC data follow closely the cavity result
for the dCP or SG phases, respectively. As expected for a first-order
transition, we find some hysteresis around $\G_{\rm c}$ before the
system finally jumps to the QP phase.  We then consider a more
interesting QMC run starting from large $\G=2$ in the QP phase and
slowly decreasing $\G$.  For both $c=3, 4$, QMC data follow the cavity
ones down to the transition, but in both cases, the energy
remains {\it extensively higher} than the ground-state energy for any
$\G < \G_{\rm c}$. This is already an important indication of the
difficulty in finding the ground state in the dCP and SG phases, even
in presence of quantum fluctuations.

\begin{figure}
  \includegraphics[width=.85\columnwidth]{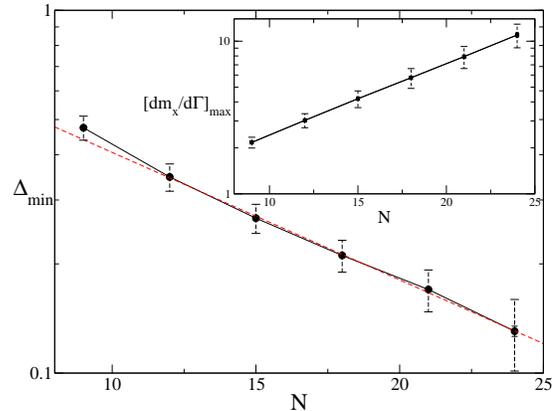}
  \vskip-10pt
  \caption{Data for $c=3$ on USA
    instances. (Main panel) Average of the minimal gap $\D_{\rm min}$ as a function of $N$. 
    Dashed line is a fit to $\D_{\rm min}(N) = 0.911
    \exp(-0.081 N)$. Inset: average of $[d m_x/d \G]_{\rm max}$.
    In both cases, error bars are of the order of the symbol size 
    except when explicitly shown ($N=24$). Dashed bars represent
    the standard deviation of a single realization of the random variables
    $\D_{\rm min}$ and $[d m_x/d \G]_{\rm max}$.
      }
  \label{fig:4}
\end{figure}

We have further investigated the consequences of the transition at
$\G_{\rm c}$ for the QAA by investigating with exact diagonalization
tools the dependence on $\G$ of the low-energy part of the spectrum of
$H$ for small sizes. To unambiguously define the gap $\D(\G)$ between the ground
state of $H$ and its first excited state at all values of $\G$, 
we concentrated on instances of $H_P$ having a
Unique Satisfying Assignment (USA), i.e., a single classical ground
state. 
For $c=3$ and $N\!\to\!\io$, the fraction of SAT and USA instances are $f_{\rm SAT} = 0.609 \pm 0.003$
and $f_{\rm USA} = 0.2850 \pm 0.0022$, as determined by 
using a Davis-Putnam-Logemann-Loveland--like algorithm to 
count the number of solutions of 40000 instances~\cite{DPLL}.
Since USA instances are a finite fraction of the uniform ensemble, they can be easily constructed.
The spectrum of a
typical USA instance of $N=15$ spins is reported in
Fig.~\ref{fig:3}. We observe, as expected, that the gap $\D(\G)$
has a minimum $\D_{\rm min}$ close to the phase transition at $\G_{\rm c}$
(recall that $\G_{\rm c} \approx 1$ for $c=3$ at $N\to\io$).  
Around the same $\G_{\rm c}$, $m_x$ changes abruptly, hence
its derivative has a large maximum $[dm_x/d\G]_{\rm max}$.
In Fig.~\ref{fig:4} we show the behavior of the average $\D_{\rm min}$ and
$[dm_x/d\G]_{\rm max}$ as a function of $N$. Our data are
clearly consistent with an exponential scaling of the gap, which is
expected in presence of a first-order transition (see~\cite{QREM} for
a discussion on how to compute the prefactor in the exponential in
fully-connected models), and an exponential divergence of $[dm_x/d\G]_{\rm max}$. 
The probability distribution over instances
of $\D_{\rm min}$ and $[dm_x/d\G]_{\rm max}$ has a unique peak close
to their average, and its variance is also reported in Fig.~\ref{fig:4} (dashed bars).
This shows that all instances undergo a first order transition of the same kind
in the thermodynamic limit.

Let us finally suggest that the main differences between our
observations and the ones of~\cite{First,Second} arises from the method of
construction of instances. Most random optimization problems that
undergo a SAT-UNSAT transition as a control parameter is continuously
tuned still have an {\it exponential} number of ground states right
before they become UNSAT.  Conditioning on USA instances is thus,
contrary to the case studied in this Letter, an exponentially rare
event which restricts the study to extremely atypical instances and
that also forbids the construction of large instances. There exists
however a natural family of difficult optimization models
(e.g., Eq.~(\ref{H}) with $c=4$ and $J_a=1 \ \forall a$) which have
unique ground states with probability 1 in the thermodynamic
limit~\cite{instances}.  This is thus a practical way of generating
large USA instances as {\it typical} ones in a uniform random
ensemble.

\paragraph*{Conclusions.---}
We have obtained the full phase diagram of the quantum regular XORSAT
optimization problem as a function of $T$ and $\G$. Our main results
are: {\it i)}~There is a first-order quantum phase transition at $T=0$
between a
Paramagnetic or a Spin-Glass phase and a Quantum Paramagnetic phase, at
a critical value of $\G=\G_{\rm c}$; {\it ii)}~The transition is due
to a crossing between the low-$\G$ classical-like ground state, and
the high-$\G$ quantum paramagnetic state. It is of very different
nature from the level crossing at infinitesimal $\G$ between different
spin-glass ground states discussed in~\cite{First}; {\it iii)}~The
first-order transition is observed for almost all instances, even for
very small $N$; {\it iv)}~The transition is associated to an
exponentially vanishing gap of $H$, hence the Quantum Adiabatic
Algorithm requires a run time scaling exponentially with system size.
These results indicate that quantum adiabatic 
computations, at least in their original formulation~\cite{annealing,adiabatic}, 
fail for difficult optimization problems.

The method introduced here is not restricted to XORSAT, and can be
applied to investigate other random optimization problems, such as, e.g., the exact cover
discussed in~\cite{Second,First,FirstY} or random $K$-SAT.  Another closely related problem is
MAP decoding of LDPC codes.
The study of these problems should be very
interesting, since in the classical case they typically have exponentially many solutions even
at the SAT-UNSAT transition. This huge degeneracy will be partially lifted
by adding a transverse field, but we expect the low energy spectrum of these
problems to be quite complicated, and characterized by many almost-degenerate
low energy states. What is the precise definition of the relevant gap for the QAA
in these cases is clearly an interesting problem. It is, however, rather
natural, based on the accumulated knowledge on classical optimization
problems~\cite{CSP_classical} and mean-field
spin-glasses~\cite{QREM,FirstMF}, to believe this first-order
transition, and thus the failure of the QAA to be a generic feature in
these problems. Yet, the instances that are encountered in practical
applications are often very different from the random instances investigated
here: the latter are characterized by a locally tree-like graph, while the former
have more structure. Extending these results to more realistic instances
is an important direction for future research.

\acknowledgments{
We wish to thank B.~Altshuler, J.~Roland and A.~P.~Young for
illuminating discussions.
}


\end{document}